\documentclass[useAMS,usenatbib,letterpaper]{mn2e}

\usepackage{times}
\usepackage{graphicx}
\usepackage{amssymb}
\usepackage[french,english]{babel}

\title[Compressed sensing for radio interferometry]
{Compressed sensing imaging techniques for radio interferometry}

\author[Wiaux et al.]
{Y. Wiaux$^{1,2}$, L. Jacques$^{1,3}$, G. Puy$^{1}$, A. M. M. Scaife$^{4}$, P. Vandergheynst$^{1}$\\
$^{1}$Institute of Electrical Engineering, Ecole Polytechnique F\'ed\'erale de Lausanne (EPFL), CH-1015 Lausanne, Switzerland\\
$^{2}$Centre for Particle Physics and Phenomenology, Universit\'e catholique de Louvain (UCL), B-1348 Louvain-la-Neuve, Belgium\\
$^{3}$Communications and Remote Sensing Laboratory, Universit\'e catholique de Louvain (UCL), B-1348 Louvain-la-Neuve, Belgium\\
$^{4}$Astrophysics Group, Cavendish Laboratory, University of Cambridge, Cambridge CB3 0HE, United Kingdom}

\begin{document}

\date{\today}

\pagerange{\pageref{firstpage}--\pageref{lastpage}} \pubyear{2009}

\maketitle

\label{firstpage}

\begin{abstract}
Radio interferometry probes astrophysical signals through incomplete
and noisy Fourier measurements. The theory of compressed sensing demonstrates
that such measurements may actually suffice for accurate reconstruction
of sparse or compressible signals. We propose new generic imaging
techniques based on convex optimization for global minimization problems
defined in this context. The versatility of the framework notably
allows introduction of specific prior information on the signals,
which offers the possibility of significant improvements of reconstruction
relative to the standard local matching pursuit algorithm CLEAN used
in radio astronomy. We illustrate the potential of the approach by
studying reconstruction performances on simulations of two different
kinds of signals observed with very generic interferometric configurations.
The first kind is an intensity field of compact astrophysical objects.
The second kind is the imprint of cosmic strings in the temperature
field of the cosmic microwave background radiation, of particular
interest for cosmology.
\end{abstract}


\section{Introduction}

\label{sec:Introduction}Radio interferometry is a powerful technique
for aperture synthesis in astronomy, dating back to more than sixty
years ago \citep{ryle46,blythe57,ryle59,ryle60,thompson04}. In a
few words, thanks to interferometric techniques, radio telescope arrays
synthesize the aperture of a unique telescope of the same size as
the maximum projected distance between two telescopes on the plane
perpendicular to the pointing direction of the instrument. This allows
observations with otherwise inaccessible angular resolutions and sensitivities
in radio astronomy. The small portion of the celestial sphere accessible
to the instrument around the pointing direction tracked during observation
defines the original real planar signal or image $I$ to be recovered.
The fundamental Nyquist-Shannon theorem requires a signal to be sampled
at a frequency of twice its bandwidth to be exactly known. The signal
$I$ may therefore be expressed as a vector $x\in\mathbb{R}^{N}$
containing the required number $N$ of sampled values. Radio-interferometric
data are acquired in the Fourier plane. The number $m$ of spatial
frequencies probed may be much smaller than the number $N$ of discrete
frequencies of the original band-limited signal, so that the Fourier
coverage is incomplete. Moreover the spatial frequencies probed are
not uniformly sampled. The measurements are also obviously affected
by noise. An ill-posed inverse problem is thus defined for reconstruction
of the original image.

Beyond the Nyquist-Shannon theorem, the emerging theory of compressed
sensing aims at merging data acquisition and compression \citep{candes06a,candes06b,candes06c,donoho06,baraniuk07}.
It notably relies on the idea that a large variety of signals in Nature
are sparse or compressible. By definition, a signal is sparse in some
basis if its expansion contains only a small number of non-zero coefficients.
More generally it is compressible if its expansion only contains a
small number of significant coefficients, i.e. if a large number of
its coefficients bear a negligible value. Compressed sensing theory
demonstrates that a much smaller number of linear measurements is
required for accurate knowledge of such signals than is required for
Nyquist-Shannon sampling. The sensing matrix must simply satisfy a
so-called restricted isometry property. In particular, a small number
of random measurements in a sensing basis incoherent with the sparsity
or compressibility basis will ensure this property with overwhelming
probability, e.g. random Fourier measurements of a signal sparse in
real or wavelet space. Consequently, if compressed sensing had been
developed before the advent of radio interferometry, one could probably
not have thought of a much better design of measurements for sparse
and compressible signals in an imaging perspective.

In this work we present results showing that the theory of compressed
sensing offers powerful image reconstruction techniques for radio-interferometric
data. These techniques are based on global minimization problems,
which are solved by convex optimization algorithms. We also emphasize
on the versatility of the scheme relative to the inclusion of specific
prior information on the signal in the minimization problems. This
versatility allows the definition of image reconstruction techniques
which are significantly more powerful than standard deconvolution
algorithm called CLEAN used in the context of radio astronomy.

In Section \ref{sec:Radio-interferometry}, we pose the inverse problem
for image reconstruction from radio-interferometric data and discuss
the standard image reconstruction techniques used in radio astronomy.
In Section \ref{sec:Compressed-sensing-perspective}, we concisely
describe the central results of the theory of compressed sensing regarding
the definition of a sensing basis and the accurate reconstruction
of sparse or compressible signals. In Section \ref{sec:Applications},
we firstly comment on the exact compliance of radio interferometric
measurements with compressed sensing. We then study the reconstruction
performances of various compressed sensing imaging techniques relative
to CLEAN on simulations of two kinds of signals of interest for astrophysics
and cosmology. We finally conclude in Section \ref{sec:Conclusion}.

Notice that a first application of compressed sensing in astronomy
\citep{bobin08} was very recently proposed for non-destructive data
compression on board the future Herschel space observatory%
\footnote{http://herschel.esac.esa.int/%
}. The versatility of the compressed sensing framework to account for
specific prior information on signals was already pointed out in that
context. Moreover, the generic potential of compressed sensing for
interferometry was pointed in the signal processing community since
the time when the theory emerged \citep{donoho06,candes06b,mary07,levanda08}.
It was also very recently acknowledged in radio astronomy \citep{cornwell08b}.
The present work nonetheless represents the first application of compressed
sensing for the definition of new imaging techniques in radio interferometry.
A huge amount of work may be envisaged along these lines, notably
for the transfer of the proposed techniques to optical and infrared
interferometry. The extension of these techniques from the plane to
the sphere will also be essential, notably with regard to forthcoming
radio interferometers with wide fields of view on the celestial sphere
\citep{cornwell08a,mcewen08}, such as the future Square Kilometer
Array (SKA)%
\footnote{http://www.skatelescope.org/%
} \citep{carilli04}.

\section{Radio interferometry}

\label{sec:Radio-interferometry}In this section, we recall the van
Cittert-Zernike theorem on the basis of which we formulate the inverse
problem posed for image reconstruction from radio-interferometric
data. We also describe and discuss the standard image reconstruction
techniques used in radio astronomy, namely a local matching pursuit
algorithm called CLEAN and a global optimization algorithm called
the maximum entropy method (MEM).

\subsection{van Cittert-Zernike theorem}

In a tracking configuration, all radio telescopes of an interferometric
array point in the same direction. The field of view observed on the
celestial sphere $\textnormal{S}^{2}$ is limited by a so-called illumination
function $A(\omega)$, depending on the angular position $\omega\in\textnormal{S}^{2}$.
The size of its angular support is essentially inversely proportional
to the size of the dishes of the telescopes \citep{thompson04}. At
each instant of observation, each telescope pair identified by an
index $b$ measures a complex visibility $y_{b}\in\mathbb{C}$. This
visibility is defined as the correlation between incoming electric
fields $E$ at the positions of the two telescopes in the three-dimensional
space, $\vec{b}_{1},\vec{b}_{2}\in\mathbb{R}^{3}$: \begin{equation}
y_{b}=\Big\langle E\left(\vec{b}_{1},t\right)E^{*}\left(\vec{b}_{2},t\right)\Big\rangle_{\Delta t}.\label{ri1}\end{equation}
In this relation, $t$ denotes the time variable and the brackets
$\langle\cdot\rangle_{\Delta t}$ denote an average over a time $\Delta t$
long relative to the period of the radio wave detected.

We consider a monochromatic signal with a wavelength of emission $\lambda$,
and made up of incoherent sources. We also consider a standard interferometer
with an illumination function whose angular support is small enough
so that the field of view may be identified to a planar patch of the
celestial sphere: $P\subset\mathbb{R}^{2}$. The signal and the illumination
function thus respectively appear as functions $I(\vec{p})$ and $A(\vec{p})$
of the angular variable seen as a two-dimensional vector $\vec{p}\in\mathbb{R}^{2}$
with an origin at the pointing direction of the array. The vector
$\vec{B}_{b}=\vec{b}_{2}-\vec{b}_{1}\in\mathbb{R}^{3}$ defining the
relative position between the two telescopes is called the baseline,
and its projection on the plane perpendicular to the pointing direction
of the instrument may be denoted as $\vec{B}_{b}^{\perp}\in\mathbb{R}^{2}$.
One also makes the additional assumption that the maximum projection
of the baselines in the pointing direction itself is small \citep{cornwell08a}.
In this context, the so-called van Cittert-Zernike theorem states
that the visibility measured identifies with the two-dimensional Fourier
transform of the image multiplied by the illumination function $AI$
at the single spatial frequency \begin{equation}
\vec{u}_{b}=\frac{\vec{B}_{b}^{\perp}}{\lambda},\label{ri2}\end{equation}
i.e. \begin{equation}
y_{b}=\widehat{AI}\left(\vec{u}_{b}\right),\label{ri3}\end{equation}
with\begin{equation}
\widehat{AI}\left(\vec{u}\right)\equiv\int_{\mathbb{R}^{2}}A\left(\vec{p}\right)I\left(\vec{p}\right)e^{-2i\pi\vec{p}\cdot\vec{u}}\textnormal{d}^{2}\vec{p},\label{ri3'}\end{equation}
for any two-dimensional vector $\vec{u}\in\mathbb{R}^{2}$. Interferometric
arrays thus probe signals at a resolution equivalent to that of a
single telescope with a size $R$ essentially equivalent to the maximum
projected baseline on the plane perpendicular to the pointing direction:
$R\simeq\max_{b}\vec{B}_{b}^{\perp}$. This expresses the essence
of aperture synthesis \citep{thompson04}.

\subsection{Interferometric inverse problem}

In the course of an observation, the projected baselines on the plane
perpendicular to the pointing direction change thanks to the Earth's
rotation and run over an ellipse in the Fourier plane of the original
image, whose parameters are linked to the parameters of observation.
The total number $m/2$ of spatial frequencies probed by all pairs
of telescopes of the array during the observation provides some Fourier
coverage characterizing the interferometer. Any interferometer is
thus simply identified by a binary mask in Fourier equal to $1$ for
each spatial frequency probed and $0$ otherwise. The visibilities
measured may be denoted as a vector of $m/2$ complex Fourier coefficients
$y\in\mathbb{C}^{m/2}=\{y_{b}=\widehat{AI}(\vec{u}_{b})\}_{1\leq b\leq m/2}$,
possibly affected by complex noise values $n\in\mathbb{C}^{m/2}=\{n_{b}=n(\vec{u}_{b})\}_{1\leq b\leq m/2}$
of astrophysical or instrumental origin. Considering that the signal
$I$ and the illumination function $A$ are real, a symmetry $\widehat{AI}(-\vec{u}_{b})=\widehat{AI}^{*}(\vec{u}_{b})$
also holds so that independent measurements may all be localized in
one half of the Fourier plane. The binary mask in Fourier identifying
the interferometer is defined in this half of the plane and rendered
symmetric around the origin so that it also corresponds to the Fourier
transform of a real function. In this context, the measured visibilities
may equivalently be denoted as a vector of $m$ real Fourier coefficients
$y\in\mathbb{R}^{m}=\{y_{r}\}_{1\leq r\leq m}$ consisting of the
real and imaginary parts of the complex measures, possibly affected
by real noise values $n\in\mathbb{R}^{m}=\{n_{r}\}_{1\leq r\leq m}$.

The original signal $I(\vec{p})$ and the illumination function $A(\vec{p})$
can be approximated by band-limited functions restricted to the finite
field of view precisely set by the illumination function: $\vec{p}\in P$.
In this context, we notice that they are identified by their Nyquist-Shannon
sampling on a discrete uniform grid of $N=N^{1/2}\times N^{1/2}$
points $\vec{p}_{i}\in\mathbb{R}^{2}$ in real space with $1\leq i\leq N$.
The sampled signal may thus be denoted as $x\in\mathbb{R}^{N}=\{x_{i}=I(\vec{p}_{i})\}_{1\leq i\leq N}$
while the illumination function is denoted as $a\in\mathbb{R}^{N}=\{a_{i}=A(\vec{p}_{i})\}_{1\leq i\leq N}$,
and the sampled product reads as $\bar{x}\in\mathbb{R}^{N}=\{\bar{x}_{i}=AI(\vec{p}_{i})\}_{1\leq i\leq N}$.
Because of the assumed finite field of view, the functions may equivalently
be described by their complex Fourier coefficients on a discrete uniform
grid of $N=N^{1/2}\times N^{1/2}$ spatial frequencies $\vec{u}_{i}$
with $1\leq i\leq N$. This grid is symmetric around the origin and
limited at the maximum frequency defining the band limit. In particular
for the Fourier coefficients of the product $AI$ one has: $\widehat{\bar{x}}\in\mathbb{C}^{N}=\{\widehat{\bar{x}}_{i}=\widehat{AI}(\vec{u}_{i})\}_{1\leq i\leq N}$.
The functions being real, one again has the symmetry $\widehat{AI}(-\vec{u}_{i})=\widehat{AI}^{*}(\vec{u}_{i})$
so that the signal is described by exactly $N/2$ complex Fourier
coefficients in one half of the Fourier plane, or equivalently $N$
real Fourier coefficients consisting of the real and imaginary parts
of these complex coefficients. In the following we only use this decomposition
with real coefficients in one half of the Fourier plane.

However the frequencies $\vec{u}_{b}$ probed defined by (\ref{ri2})
for $1\leq b\leq m/2$ are continuous and do not generally belong
to the set of discrete frequencies $\vec{u}_{i}$ for $1\leq i\leq N$.
Reconstruction schemes in general perform a preliminary gridding operation
on the visibilities $y_{r}$ with $1\leq r\leq m$ so that the inverse
problem may be reformulated in a pure discrete setting, i.e. between
the discrete Fourier and real planes \citep{thompson04}. The essential
reason for the gridding resides in the subsequent use of the standard
fast Fourier transform (FFT)%
\footnote{Notice that fast algorithms have been developed to compute a Fourier
transform on non-equispaced spatial frequencies (NFFT) \citep{potts01}.
This could in principle allow one to avoid an explicit gridding operation.%
}. For the sake of the considerations that follow we assume that the
frequencies probed $\vec{u}_{b}$ belong to the discrete grid of points
$\vec{u}_{i}$ so that no artifact due to the gridding is introduced.
In this discrete setting the Fourier coverage is unavoidably incomplete
in the sense that the number of real constraints $m$ is always smaller
than the number of unknowns $N$: $m<N$. An ill-posed inverse problem
is thus defined for the reconstruction of the signal $x$ from the
measured visibilities $y$ as: \begin{equation}
y=\Phi_{_{\!{\rm ri}}}x+n,\label{ri4}\end{equation}
for a given noise $n$, and with a sensing matrix $\Phi_{_{\!{\rm ri}}}$
for radio interferometry of the form \begin{equation}
\Phi_{_{\!{\rm ri}}}=MFD.\label{ri5}\end{equation}
In this relation, the matrix $D\in\mathbb{R}^{N\times N}=\{D_{ii'}=a_{i}\delta_{ii'}\}_{1\leq i,i'\leq N}$
is the diagonal matrix implementing the illumination function, and
the matrix $F\in\mathbb{R}^{N\times N}=\{F_{ii'}\}_{1\leq i,i'\leq N}$
implements the discrete Fourier transform providing the real Fourier
coefficients in one half of the Fourier plane. The matrix $M\in\mathbb{R}^{m\times N}=\{M_{ri}\}_{1\leq r\leq m;1\leq i\leq N}$
is the rectangular binary matrix implementing the mask characterizing
the interferometer in one half of the Fourier plane. It contains only
one non-zero value on each line, at the index of one of the two real
Fourier coefficients corresponding to each of the spatial frequencies
probed.

We restrict our considerations to independent Gaussian noise with
variance $\sigma_{r}^{2}=\sigma^{2}(y_{r})$. From a statistical point
of view, the likelihood $\mathcal{L}$ associated with a candidate
reconstruction $x^{*}$ of the signal $x$ is defined as the probability
of the data $y$ given the model $x^{*}$, or equivalently the probability
of the noise residual $n^{*}=y-\Phi_{_{\!{\rm ri}}}x^{*}$. Under
the Gaussian noise assumption it reads as\begin{equation}
\mathcal{L}\left(y\vert x^{*}\right)\propto\exp\left[-\frac{1}{2}\chi^{2}\left(x^{*};\Phi_{_{\!{\rm ri}}},y\right)\right],\label{ri6}\end{equation}
with the corresponding negative logarithm\begin{equation}
\chi^{2}\left(x^{*};\Phi_{_{\!{\rm ri}}},y\right)=\sum_{r=1}^{m}\frac{\left(n_{r}^{*}\right)^{2}}{\sigma_{r}^{2}},\label{ri7}\end{equation}
following a chi-square distribution with $m$ degrees of freedom.
The $\chi^{2}$ defines a noise level estimator. The level of residual
noise $n^{*}$ should be reduced by finding $x^{*}$ minimizing this
$\chi^{2}$, which corresponds to maximize the likelihood $\mathcal{L}$.
Typically, the measurement constraint on the reconstruction may be
defined as a bound\begin{equation}
\chi^{2}\left(x^{*};\Phi_{_{\!{\rm ri}}},y\right)\leq\epsilon^{2},\label{ri8}\end{equation}
with $\epsilon^{2}$ corresponding to some $(100\alpha)^{\textnormal{th}}$
percentile of the chi-square distribution, i.e. $p(\chi^{2}\leq\epsilon^{2})=\alpha$
for some $\alpha\leq1$. For a solution with a $\chi^{2}=\epsilon^{2}$,
there is a probability $\alpha$ that pure noise gives a residual
smaller than or equal to the observed residual $n^{*}$, and a probability
$1-\alpha$ that noise gives a larger residual. Too small an $\alpha$
would thus induce possible noise over-fitting, i.e. inclusion of part
of the noise in the reconstruction. These considerations might of
course be generalized to other kinds of noise distributions.

The inverse problem being ill-posed, many signals may formally satisfy
measurement constraints such as (\ref{ri8}). In general, the problem
may only find a unique solution $x^{*}$, as close as possible to
the true signal $x$, through a regularization scheme which should
encompass enough prior information on the original signal. All possible
image reconstruction algorithms will essentially be distinguished
through the kind of regularization considered.

\subsection{Standard imaging techniques}

The general inverse problem (\ref{ri4}) is to be considered if one
wishes to undo the multiplication by the illumination function and
to recover the original signal $x$ on the given field of view. In
practice, the reconstruction is usually considered for the original
image $I$ already multiplied by the illumination function $A$, whose
sampled values are $\bar{x}=Dx\in\mathbb{R}^{N}=\{\bar{x}_{i}=a_{i}x_{i}\}_{1\leq i\leq N}$.
In this setting the inverse problem reads as \begin{equation}
y=\bar{\Phi}_{_{\!{\rm ri}}}\bar{x}+n,\label{eq:ri4'}\end{equation}
with a sensing matrix $\bar{\Phi}_{_{\!{\rm ri}}}$ strictly implementing
a convolution: \begin{equation}
\bar{\Phi}_{_{\!{\rm ri}}}=MF.\label{eq:ri5'}\end{equation}

Firstly, the most standard and otherwise already very effective image
reconstruction algorithm from visibility measurements is called CLEAN.
It approaches the image reconstruction in terms of the corresponding
deconvolution problem in real space \citep{hogbom74,schwarz78,thompson04}.
In standard vocabulary, the inverse transform of the Fourier measurements
with all non-observed visibilities set to zero is called the dirty
image. Its sampled values $\bar{x}^{(d)}\in\mathbb{R}^{N}=\{\bar{x}_{i}^{(d)}\}_{1\leq i\leq N}$
are simply obtained by application of the adjoint sensing matrix to
the observed visibilities: $\bar{x}^{(d)}=\bar{\Phi}_{_{\!{\rm ri}}}^{\dagger}y$.
The inverse transform of the binary mask identifying the interferometer
is called the dirty beam. Its sampled values $d\in\mathbb{R}^{N}=\{d_{i}\}_{1\leq i\leq N}$
follow from the application of the adjoint sensing matrix to a vector
of unit values $1_{m}\in\mathbb{R}^{m}$: $d=\bar{\Phi}_{_{\!{\rm ri}}}^{\dagger}1_{m}$.
The inverse transform of the noise $n$ with all non-observed visibilities
set to zero defines an alternative expression of the noise in real
space. Again its sampled values $n^{(d)}\in\mathbb{R}^{N}=\{n_{i}^{(d)}\}_{1\leq i\leq N}$
are simply obtained by application of the adjoint sensing matrix to
the noise realization: $n^{(d)}=\bar{\Phi}_{_{\!{\rm ri}}}^{\dagger}n$.
The inverse problem (\ref{eq:ri4'}) can thus be rephrased by expressing
the dirty image as the convolution of the original image with the
dirty beam, plus the noise:

\begin{equation}
\bar{x}^{(d)}=d\star\bar{x}+n^{(d)}.\label{eq:clean}\end{equation}
CLEAN is a non-linear deconvolution method relying on this relation
and working by local iterative beam removal. At each iteration, the
point in real space is identified where a residual image, initialized
to the dirty image, takes its maximum absolute value. The beam is
removed at that point with the correct amplitude to produce the residual
image for the next iteration. Simultaneously the maximum absolute
value observed renormalized by the central value of the beam is added
at the same point in the approximation image, initialized to a null
image. This procedure assumes that the original signal is a sum of
Dirac spikes. A sparsity or compressibility prior on the original
signal in real space is implicitly introduced so that its energy is
concentrated at specific locations. On the contrary, the Gaussian
noise should be distributed everywhere on the image and should not
significantly affect the selection of points in the iterations. This
underlying sparsity hypothesis serves as a regularization of the inverse
problem.

A loop gain factor $\gamma$ is generally introduced in the procedure
which defines the fraction of the beam considered at each iteration.
Values $\gamma$ around a few tenths are usually used which allow
for a more cautious consideration of the sidelobes of the dirty beam.
The overall procedure is greatly enhanced by this simple improvement,
albeit at high computational cost. In a statistical sense, the stopping
criterion for the iteration procedure should be set in terms of relation
(\ref{ri8}). However, the procedure is known to be slow and the algorithm
is often stopped after an arbitrary number of iterations.

Various weighting schemes can be applied to the binary mask in Fourier.
Natural weighting simply corresponds to replace the unit values by
the inverse variance of the noise affecting the corresponding visibility
measurement. This corresponds to a standard matched filtering operation
allowing the maximization of the signal-to-noise ratio of the dirty
image before deconvolution. So-called uniform and robust weightings
can notably be used to correct for the non-uniformity of the Fourier
coverage associated with the measured visibilities and to reduce the
sidelobes of the dirty beam in real space. Multi-scale versions of
this method were also developed \citep{cornwell08b}.

CLEAN and multi-scale versions may actually be formulated in terms
of the well-known matching pursuit (MP) procedure \citep{mallat93,mallat98}.
The corresponding MP algorithm simply uses a circulant dictionary
for which the projection on atoms corresponds to the convolution with
the dirty beam. The loop gain factor may also be trivially introduced
in this context.

Secondly, another approach to the reconstruction of images from visibility
measurements is MEM. In contrast to CLEAN, MEM solves a global optimization
problem in which the inverse problem (\ref{eq:ri4'}) is regularized
by the introduction of an entropic prior on the signal \citep{ables74,gull78,cornwell85,gull99}.
For positive signals, the relative entropy function between a sampled
signal $\bar{x}\in\mathbb{R}^{N}=\{\bar{x}_{i}\}_{1\leq i\leq N}$
and a model $z\in\mathbb{R}^{N}=\{z_{i}\}_{1\leq i\leq N}$ takes
the simple form\begin{equation}
S\left(\bar{x},z\right)=-\sum_{i}^{N}\bar{x}_{i}\ln\frac{\bar{x}_{i}}{z_{i}}.\label{ri9}\end{equation}
This function is always negative and takes its maximum null value
when $\bar{x}=z$. In the absence of a precise knowledge of the signal
$\bar{x}$, $z$ is set to a vector of constant values. In such a
case, maximizing the entropy prior promotes smoothness of the reconstructed
image.

The MEM problem is the unconstrained optimization problem defined
as the minimization of a functional corresponding to the sum of the
relative entropy $S$ and the $\chi^{2}$:\begin{equation}
\min_{\bar{x}'\in\mathbb{R}^{N}}\left[\frac{1}{2}\chi^{2}\left(\bar{x}';\bar{\Phi}_{_{\!{\rm ri}}},y\right)-\tau S\left(\bar{x}',z\right)\right],\label{eq:mem}\end{equation}
for some suitably chosen regularization parameter $\tau>0$. In general,
the minimization thus requires a trade-off between $\chi^{2}$ minimization,
and relative entropy maximization.

Notice that the definition (\ref{ri9}) may easily be generalized
for non-positive signals. A multi-scale version of MEM was also defined.
It considers that the original image may have an efficient representation
in terms of its decomposition in a wavelet basis. The entropy is then
defined directly on the wavelet coefficients of the signal \citep{maisinger04}.

For completeness we finally quote the WIPE reconstruction procedure
which also solves a global minimization problem, but in which the
inverse problem (\ref{eq:ri4'}) is regularized by the introduction
of a smoothness prior on the part of the signal whose Fourier support
corresponds to the non-probed spatial frequencies. This corresponds
to minimize the $\chi^{2}$ after assigning a null value to all initially
non-observed visibilities \citep{lannes94,lannes96}.

In conclusion, CLEAN is a local iterative deconvolution technique,
while MEM and WIPE are reconstruction techniques based on global minimization
problems. All three approaches are flexible enough to consider various
bases (Dirac, wavelet, etc.) where a majority of natural signals can
have a sparse or compressible representation. CLEAN also implicitly
assumes the sparsity of the signal in the reconstruction procedure.
But none of these methods explicitly imposes the sparsity or compressibility
prior on the reconstruction. This precise gap is notably bridged by
the imaging techniques defined in the framework of the compressed
sensing theory.

\section{Compressed sensing}

\label{sec:Compressed-sensing-perspective}In this section we define
the general framework of the theory of compressed sensing and quote
its essential impact beyond the Nyquist-Shannon sampling theorem.
We then describe the restricted isometry property that the sensing
basis needs to satisfy so that sparse and compressible signals may
be accurately recovered through a global optimization problem. We
finally discuss the idea that incoherence of the sensing and sparsity
or compressibility bases as well as randomness of the measurements
are the key properties to ensure this restricted isometry.

\subsection{Beyond Nyquist-Shannon}

In the framework of compressed sensing the signals probed are firstly
assumed to be sparse or compressible in some basis. Technically, we
consider a real signal identified by its Nyquist-Shannon sampling
as $x\in\mathbb{R}^{N}=\{x_{i}\}_{1\leq i\leq N}$. A real basis $\Psi\in\mathbb{R}^{N\times T}=\{\Psi_{iw}\}_{1\leq i\leq N;1\leq w\leq T}$
is defined, which may be either orthogonal, with $T=N$, or redundant,
with $T>N$ \citep{rauhut08}. The decomposition $\alpha\in\mathbb{R}^{T}=\{\alpha_{w}\}_{1\leq w\leq T}$
of the signal defined by\begin{equation}
x=\Psi\alpha,\label{csp1}\end{equation}
is sparse or compressible in the sense that it only contains a small
number $K\ll N$ of non-zero or significant coefficients respectively.
The signal is then assumed to be probed by $m$ real linear measurements
$y\in\mathbb{R}^{m}=\{y_{r}\}_{1\leq r\leq m}$ in some real sensing
basis $\Phi\in\mathbb{R}^{m\times N}=\{\Phi_{ri}\}_{1\leq r\leq m;1\leq i\leq N}$
and possibly affected by independent and identically distributed noise
$n\in\mathbb{R}^{m}=\{n_{r}\}_{1\leq r\leq m}$: \begin{equation}
y=\Theta\alpha+n\textnormal{ with }\Theta=\Phi\Psi\in\mathbb{R}^{m\times T}.\label{csp2}\end{equation}
This number $m$ of constraints is typically assumed to be smaller
than the dimension $N$ of the vector defining the signal, so that
the inverse problem (\ref{csp2}) is ill-posed.

In this context, the theory of compressed sensing defines the explicit
restricted isometry property (RIP) that the matrix $\Theta$ should
satisfy in order to allow an accurate recovery of sparse or compressible
signals \citep{candes06a,candes06b,candes06c}. In that regard, the
theory offers multiple ways to design suitable sensing matrices $\Phi$
from properties of incoherence with $\Psi$ and randomness of the
measurements. It shows in particular that a small number of measurements
is required relative to a naive Nyquist-Shannon sampling: $m\ll N$.
The framework also defines a global minimization problem for the signal
recovery called Basis Pursuit (BP). This problem regularizes the originally
ill-posed inverse problem by an explicit sparsity or compressibility
prior on the signal. The corresponding solution may be obtained through
convex optimization. Alternative global minimization problems may
also be designed.

\subsection{Restricted isometry and Basis Pursuit}

Let us primarily recall that the $\ell_{p}$ norm of a real vector
$u\in\mathbb{C}^{Q}=\{u_{l}\}_{1\leq l\leq Q}$ is defined for any
$p\in\mathbb{R}_{+}$ as $\vert\vert u\vert\vert_{p}\equiv(\sum_{l=1}^{Q}\vert u_{l}\vert^{p})^{1/p}$,
where $\vert u_{l}\vert$ stands for the absolute value of the component
$u_{l}$. The well-known $\ell_{2}$ norm is to the square-root of
the sum of the absolute values squared of the vector components.

By definition the matrix $\Theta$ satisfies a RIP of order $K$ if
there exists a constant $\delta_{K}<1$ such that\begin{equation}
\left(1-\delta_{K}\right)\vert\vert\alpha_{K}\vert\vert_{2}^{2}\leq\vert\vert\Theta\alpha_{K}\vert\vert_{2}^{2}\leq\left(1-\delta_{K}\right)\vert\vert\alpha_{K}\vert\vert_{2}^{2},\label{csp3}\end{equation}
for all vectors $\alpha_{K}$ containing at maximum $K$ non-zero
coefficients.

The $\ell_{1}$ norm of the vector $\alpha\in\mathbb{R}^{T}=\{\alpha_{w}\}_{1\leq w\leq T}$
is simply defined as the sum of the absolute values of the vector
components:\begin{equation}
\vert\vert\alpha\vert\vert_{1}\equiv\sum_{w=1}^{T}\vert\alpha_{w}\vert.\label{csp4}\end{equation}
From a Bayesian point of view, this $\ell_{1}$ norm may be seen as
the negative logarithm of a Laplacian prior distribution on each independent
component of $\alpha$. For comparison the square of the $\ell_{2}$
norm may be seen as the negative logarithm of a Gaussian prior distribution.
It is well-known that a Laplacian distribution is highly peaked and
bears heavy tails, relative to a Gaussian distribution. This corresponds
to say that the signal is defined by only a small number of significant
coefficients, much smaller than a Gaussian signal would be. In other
words the representation $\alpha$ of the signal $x$ in the sparsity
or compressibility basis $\Psi$ is indeed sparse or compressible
if it follows such a prior. Finding the $\alpha'$ that best corresponds
to this prior requires to maximize its Laplacian probability distribution,
or equivalently to minimize the $\ell_{1}$ norm. Notice that this
conclusion also follows from a pure geometrical argument in $\mathbb{R}^{T}$
\citep{candes06b,baraniuk07}.

A constrained optimization problem explicitly regularized by a $\ell_{1}$
sparsity prior can be defined. This so-called Basis Pursuit denoise
($\textnormal{BP}_{\epsilon}$) problem is the minimization of the
$\ell_{1}$ norm of $\alpha'$ under a constraint on the $\ell_{2}$
norm of the residual noise:\begin{equation}
\min_{\alpha'\in\mathbb{R}^{T}}\vert\vert\alpha'\vert\vert_{1}\textnormal{ subject to }\vert\vert y-\Theta\alpha'\vert\vert_{2}\leq\epsilon.\label{eq:bpepsilon}\end{equation}
Let us recall that the noise was assumed to be identically distributed.
Consequently, considering Gaussian noise, the $\ell_{2}$ norm term
in the $\textnormal{BP}_{\epsilon}$ problem is identical to the condition
(\ref{ri8}), for $\epsilon^{2}$ corresponding to some suitable percentile
of the $\chi^{2}$ distribution with $m$ degrees of freedom governing
the noise level estimator. This $\textnormal{BP}_{\epsilon}$ problem
is solved by application of non-linear and iterative convex optimization
algorithms \citep{combettes07,vandenBerg08}. In the absence of noise,
the $\textnormal{BP}_{\epsilon}$ problem is simply called Basis Pursuit
(BP). If the solution of the $\textnormal{BP}_{\epsilon}$ problem
is denoted $\alpha^{*}$ then the corresponding synthesis-based signal
reconstruction reads, from (\ref{csp1}), as $x^{*}=\Psi\alpha^{*}$.

Compressed sensing shows that if the matrix $\Theta$ satisfies a
RIP of order $2K$ with some suitable constant $\delta_{2K}<\sqrt{2}-1$
\citep{candes08b}, then the solution $x^{*}$ of the $\textnormal{BP}_{\epsilon}$
problem provides an accurate reconstruction of a signal $x$ that
is sparse or compressible with $K$ significant coefficients. The
reconstruction may be said to be optimal in that exactly sparse signals
are recovered exactly through BP in the absence of noise: $x^{*}=x$.
Moreover strong stability results exist for compressible signals in
the presence of noise. In that case, the $\ell_{2}$ norm of the difference
between the representation $\alpha$ of the signal in the sparsity
or compressibility basis and its reconstruction $\alpha^{*}$ is bounded
by the sum of two terms. The first term is due to the noise and is
proportional to $\epsilon$. The second term is due to the non-exact
sparsity of a compressible signal and is proportional to the $\ell_{1}$
norm of the difference between $\alpha$ and the approximation $\alpha_{K}$
defined by retaining only its $K$ largest components and sending
all other values to zero. In this context, one has \begin{equation}
\vert\vert\alpha-\alpha^{*}\vert\vert_{2}\leq C_{1,K}\epsilon+C_{2,K}\frac{\vert\vert\alpha-\alpha_{K}\vert\vert_{1}}{\sqrt{K}},\label{csp5}\end{equation}
for two known constants $C_{1,K}$ and $C_{2,K}$ depending on $\delta_{2K}$.
For instance, when $\delta_{2K}=0.2$, we have $C_{1,K}=8.5$ and
$C_{2,K}=4.2$ \citep{candes06b,candes08b}. In an orthonormal basis
$\Psi$ this relation represents an explicit bound on the $\ell_{2}$
norm of the difference between the signal $x$ itself and its reconstruction
$x^{*}$ as $\vert\vert x-x^{*}\vert\vert_{2}=\vert\vert\alpha-\alpha^{*}\vert\vert_{2}$.
Moreover $x_{K}=\Psi\alpha_{K}$ then represents the best sparse approximation
of $x$ with $K$ terms, in the sense that $\vert\vert x-x_{K}\vert\vert_{2}$
is minimum.

The constrained $\textnormal{BP}_{\epsilon}$ problem may also be
rephrased in terms of an unconstrained minimization problem for a
functional defined as the sum of the $\ell_{1}$ norm of $\alpha'$
and the $\ell_{2}$ norm of the residual noise:\begin{equation}
\min_{\alpha'\in\mathbb{R}^{T}}\left[\frac{1}{2}\vert\vert y-\Theta\alpha'\vert\vert_{2}^{2}+\tau\vert\vert\alpha'\vert\vert_{1}\right],\label{csp6}\end{equation}
for some suitably chosen regularization parameter $\tau>0$. For each
value of $\epsilon$, there exists a value $\tau$ such that the solutions
of the constrained and unconstrained $\ell_{1}$ sparsity problems
are identical \citep{vandenBerg08}. From a Bayesian point of view,
this minimization is then equivalent to maximum a posteriori (MAP)
estimation for a signal with Laplacian prior distribution in the sparsity
or compressibility basis, in the presence of Gaussian noise.

Finally, alternative minimization problems may be defined for the
recovery. Firstly, a $\ell_{p}$ norm with $0<p\leq1$ may for example
be substituted for the $\ell_{1}$ norm in the definition of the minimization
problem. From a Bayesian point of view, the $\ell_{p}$ norm to the
power $p$ may be seen as the negative logarithm of a prior distribution
identified as a generalized Gaussian distribution (GGD). Such distributions
are even more highly peaked and bear heavier tails than a Laplacian
distribution and thus promote stronger compressibility of the signals.
Theoretical results hold for such $\ell_{p}$ norm minimization problems
when a RIP is satisfied \citep{foucart08}. Such problems are non-convex
but can be solved iteratively by convex optimization algorithms performing
re-weighted $\ell_{1}$ norm minimization \citep{candes08a,davies08,foucart08,chartrand07}.
Secondly, a TV norm may also be substituted for the $\ell_{1}$ norm
in the definition of the minimization problem for signals with sparse
or compressible gradients. The TV norm of a signal is simply defined
as the $\ell_{1}$ norm of the magnitude of its gradient \citep{rudin92}.
A theoretical result of exact reconstruction holds for such TV norm
minimization problems in the case of Fourier measurements of signals
with exactly sparse gradients in the absence of noise \citep{candes06a}.
But no proof of stability relative to noise and non-exact sparsity
exists at the moment. Such minimization is also accessible through
an iterative scheme from convex optimization algorithms \citep{candes05}.

This flexibility in the definition of the optimization problem is
a first important manifestation of the versatility of the compressed
sensing theory, and of the convex optimization scheme. It opens the
door to the definition a whole variety of powerful image reconstruction
techniques that may take advantage of some available specific prior
information on the signal under scrutiny beyond its generic sparsity
or compressibility.

\subsection{Incoherence and randomness}

The issue of the design of the sensing matrix $\Phi$ ensuring the
RIP for $\Theta=\Phi\Psi$ is of course fundamental. One can actually
show that incoherence of $\Phi$ with the sparsity or compressibility
basis $\Psi$ and randomness of the measurements will ensure that
the RIP is satisfied with overwhelming probability, provided that
the number of measurements is large enough relative to the sparsity
$K$ considered \citep{candes06b,candes06c}. In this context, the
variety of approaches to design suitable sensing matrices is a second
form of the versatility of the compressed sensing framework. 

As a first example, the measurements may be drawn from a Gaussian
matrix $\Phi$ with purely random real entries, in which case the
RIP is satisfied if \begin{equation}
K\leq\frac{Cm}{\ln(N/m)},\label{csp7}\end{equation}
for some constant $C$. The most recent result provides a value $C\simeq0.5$,
hence showing that the required redundancy of measurements $m/K$
is very small \citep{donoho09}.

As a second example of interest for radio interferometry, the measurements
may arise from a uniform random selection of Fourier frequencies.
In this case, the precise condition for the RIP depends on the degree
of incoherence between the Fourier basis and the sparsity or compressibility
basis. If the unit-normed basis vectors corresponding to the lines
of $F$ and the columns of $\Psi$ are denoted $\{f_{e}\}_{1\leq e\leq N}$
and $\{\psi_{e'}\}_{1\leq e'\leq T}$, the mutual coherence $\mu$
of the bases may be defined as their maximum scalar product: \begin{equation}
\mu=\sqrt{N}\max_{e,e'}\vert\langle f_{e}\vert\psi_{e'}\rangle\vert.\label{csp8}\end{equation}
The RIP is then satisfied if \begin{equation}
K\leq\frac{C'm}{\mu^{2}\ln^{4}N},\label{csp9}\end{equation}
for some constant $C'$. As the incoherence is maximum between the
Fourier and real spaces with $\mu=1$, the lowest number of measurements
would be required for a signal that is sparse in real space. Notice
that a factor $\ln N$ instead of $\ln^{4}N$ in condition (\ref{csp9})
was not proven but conjectured, suggesting that a lower number of
measurements would still ensure the RIP. In that regard, empirical
results \citep{lustig07} suggest that ratios $m/K$ between $3$
and $5$ already ensure a reconstruction quality through $\textnormal{BP}_{\epsilon}$
that is equivalent to the quality ensured by (\ref{csp5}).

Let us also emphasize that the TV norm minimization is often used
from Fourier measurements of signals with sparse or compressible gradients.
As already stated no stability result such as (\ref{csp5}) was proven
for the reconstruction provided by this minimization scheme. Empirical
results suggest however that TV norm minimization provides the same
quality of reconstruction as $\textnormal{BP}_{\epsilon}$ for the
same typical ratios $m/K$ between $3$ and $5$ \citep{candes05,lustig07}.

\section{Applications}

\label{sec:Applications}In this section, we firstly comment on the
exact compliance of radio interferometric measurements with compressed
sensing. We then consider simulations of two kinds of signals for
reconstruction from visibility measurements: an intensity field of
compact astrophysical objects and a signal induced by cosmic strings
in the temperature field of the cosmic microwave background (CMB)
radiation. Relying on the versatility of the convex optimization scheme,
enhanced minimization problems are defined in the compressed sensing
perspective through the introduction of specific prior information
on the signals. The reconstruction performance is studied in comparison
both with the standard $\textnormal{BP}_{\epsilon}$ reconstructions
in the absence of specific priors and with the CLEAN reconstruction.

\subsection{Interferometric measurements and compressed sensing}

In the context of compressed sensing, the sensing matrix needs to
satisfy the RIP. If Fourier measurements are considered, this requirement
may be reached through a uniform random selection of a low number
of Fourier frequencies. In the context of radio interferometry, realistic
visibility distributions are deterministic, i.e. non-random, superpositions
of elliptical distributions in the Fourier plane of the image to reconstruct.
However, the structure of the Fourier sampling is extremely dependent
on the specific configuration of the radio telescope array under consideration.
Visibilities from various interferometers may be combined, as well
as visibilities from the same interferometer with different pointing
directions in the mosaicking technique \citep{thompson04}. From this
point of view the realistic visibility distributions themselves are
rather flexible. Moreover, the standard uniform weighting of the visibilities
may be used to provide uniformity of the effective measurement density
in the Fourier plane. Correctly studied realistic distributions might
thus not be so far from complying exactly with the compressed sensing
requirements. Finally, it was recently suggested that specific deterministic
distributions of a low number of linear measurements might in fact
allow accurate signal reconstruction in the context of compressed
sensing \citep{matei08}.

Nonetheless, modifications of radio interferometric measurements might
be conceived in order to comply exactly with standard compressed sensing
results. To this end, one might want to introduce randomness in the
visibility distribution. Formally, random repositioning of the telescopes
during observation or random integration times for the definition
of individual visibilities could provide important advances in that
direction. Also notice that compressed sensing does not require that
measurements be identified to Fourier coefficients of the signal.
The versatility of the framework relative to the design of suitable
sensing matrices might actually be used to define generalized radio
interferometric measurements, beyond standard visibilities, ensuring
that the RIP is explicitly satisfied. In this perspective, direct
modifications of the acquisition process through a scheme similar
to spread spectrum techniques \citep{naini09} or coded aperture techniques
\citep{marcia08} could also provide important advances.

In the following applications we simply consider standard visibility
measurements. We assume generic interferometric configurations characterized
by uniform random selections of visibilities.

\subsection{Experimental set up}

\begin{figure*}
\begin{center}
\includegraphics[height=4cm,keepaspectratio]{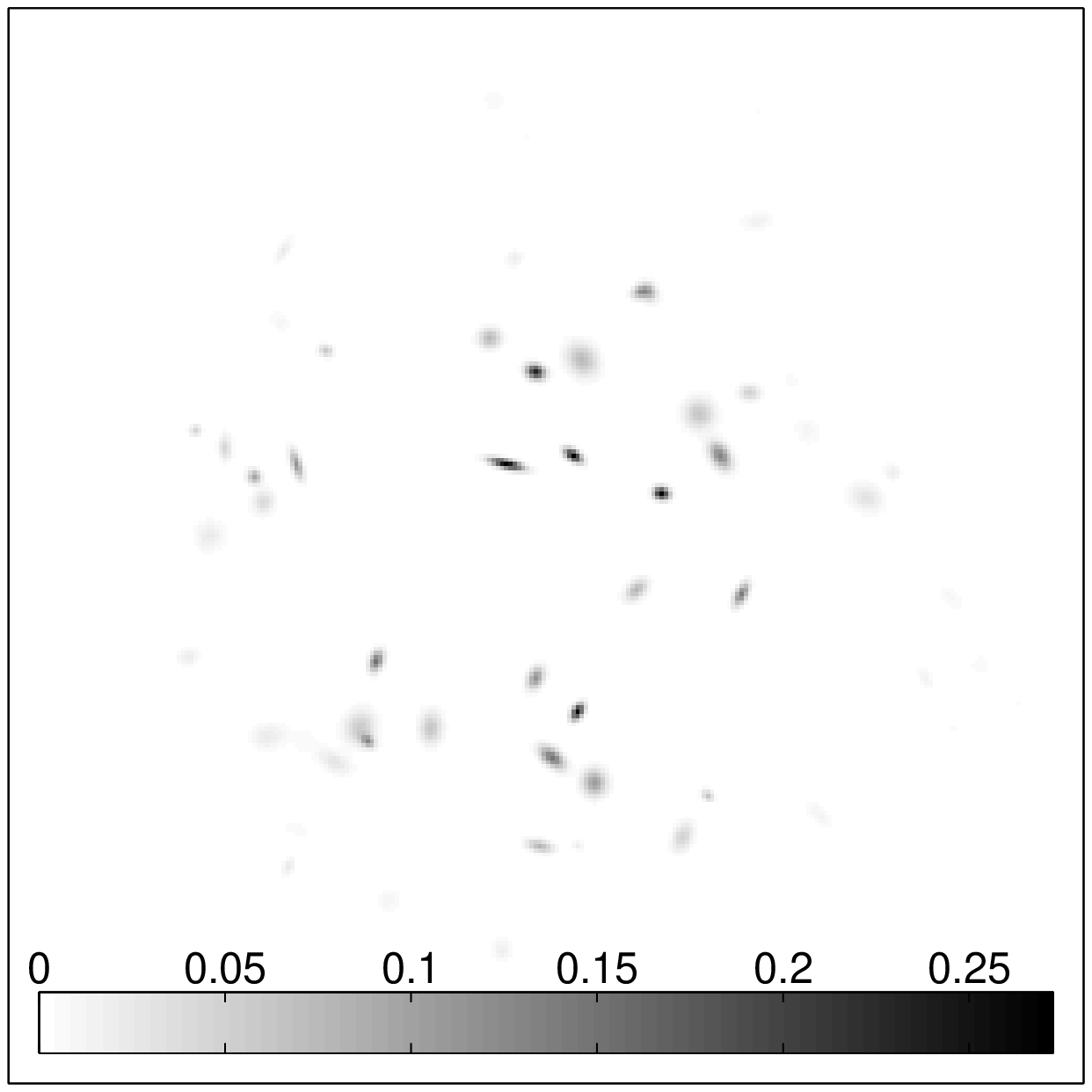}
\hspace{0.2cm}
\raisebox{-0.7mm}{\includegraphics[height=4.07cm,keepaspectratio]{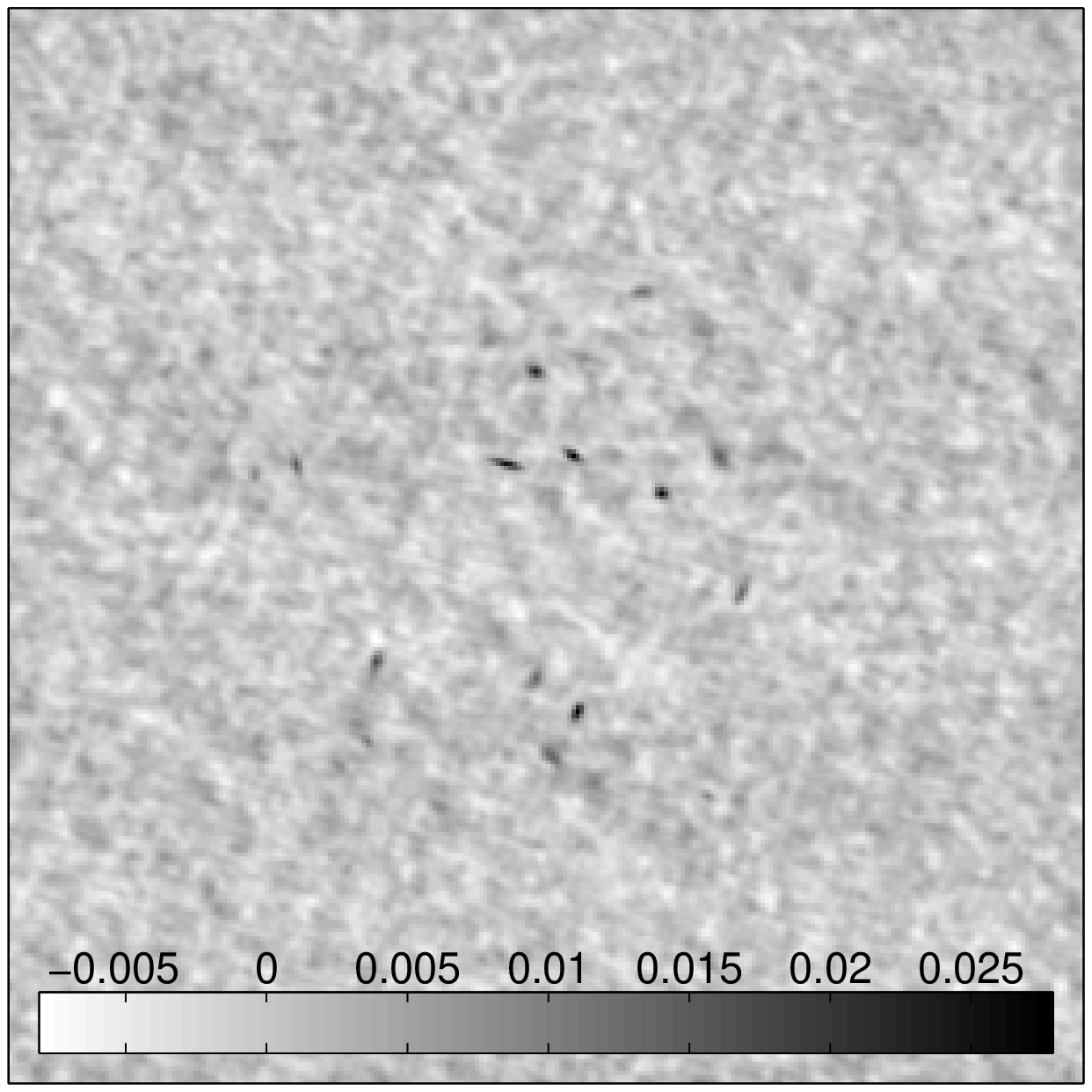}}
\hspace{0.2cm}
\includegraphics[height=4cm,keepaspectratio]{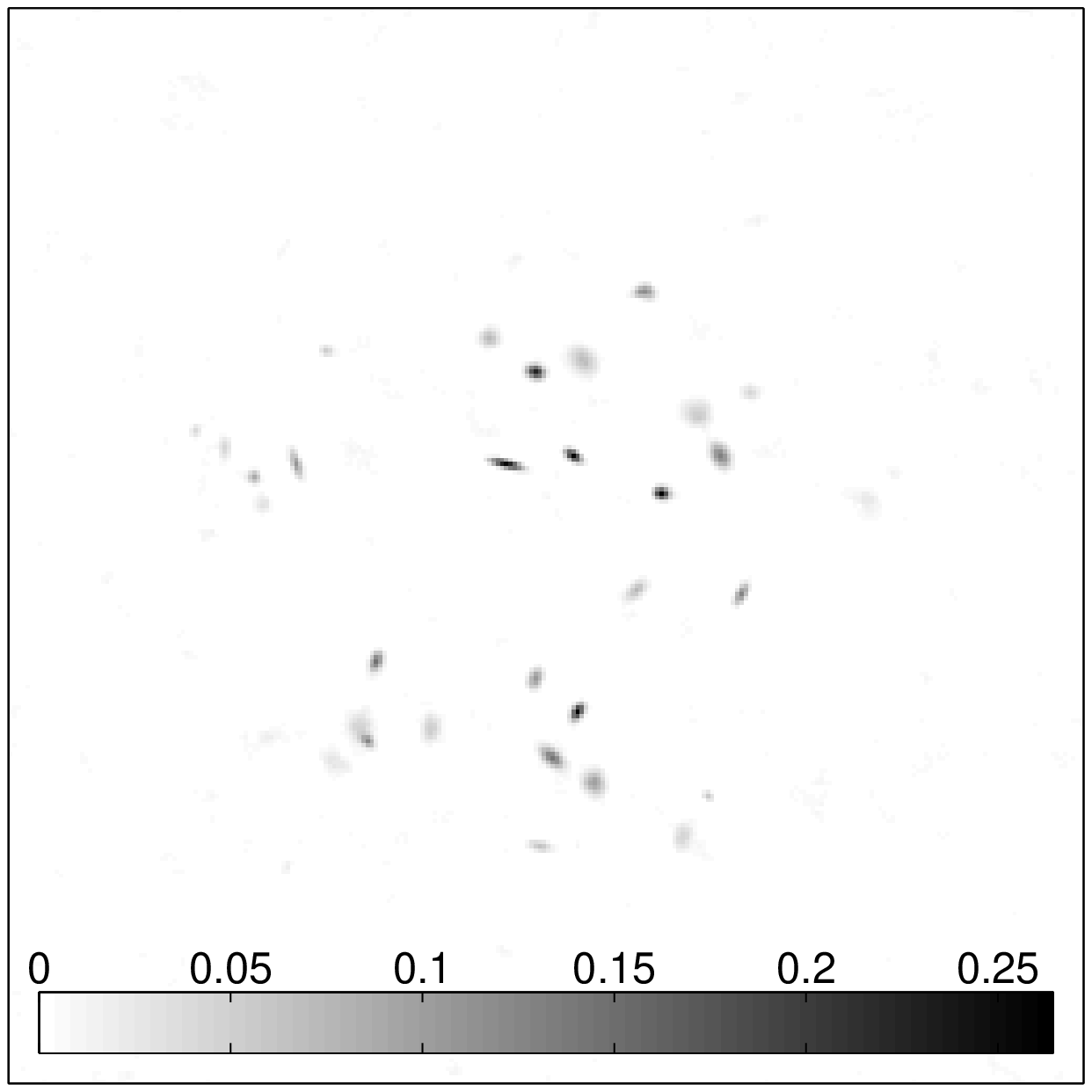}
\hspace{0.2cm}
\includegraphics[height=4cm,keepaspectratio]{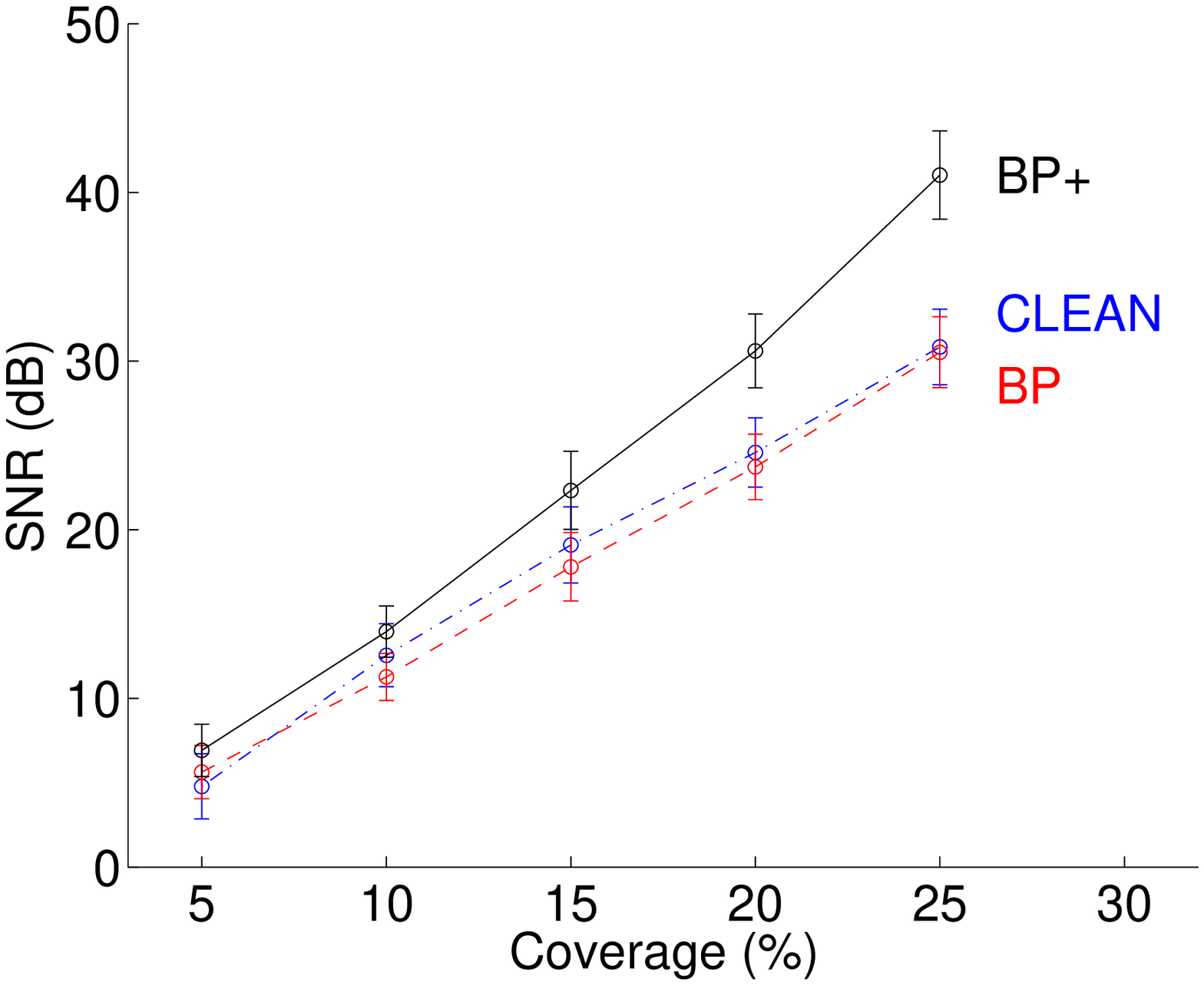}
\includegraphics[height=4cm,keepaspectratio]{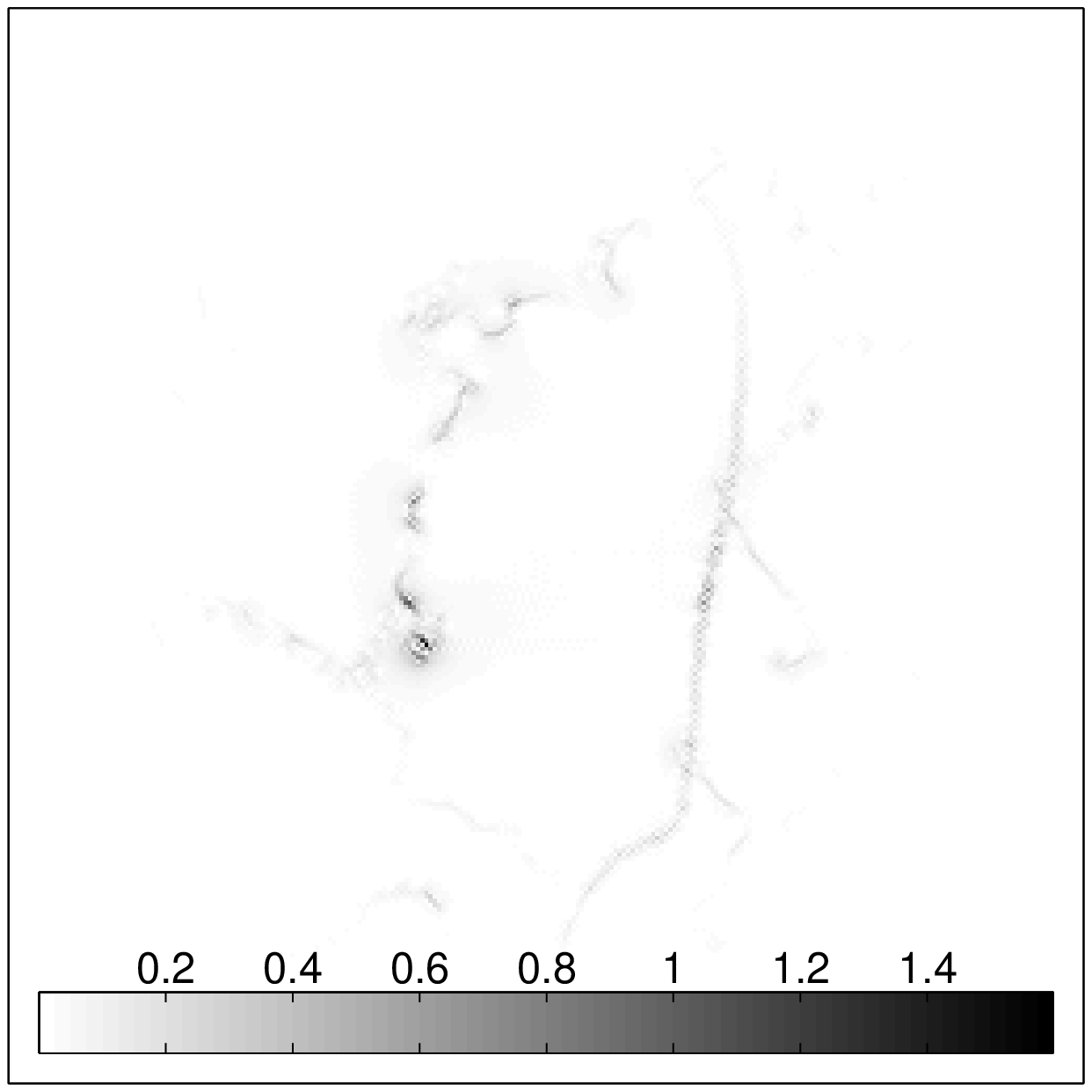}
\hspace{0.2cm}
\includegraphics[height=4cm,keepaspectratio]{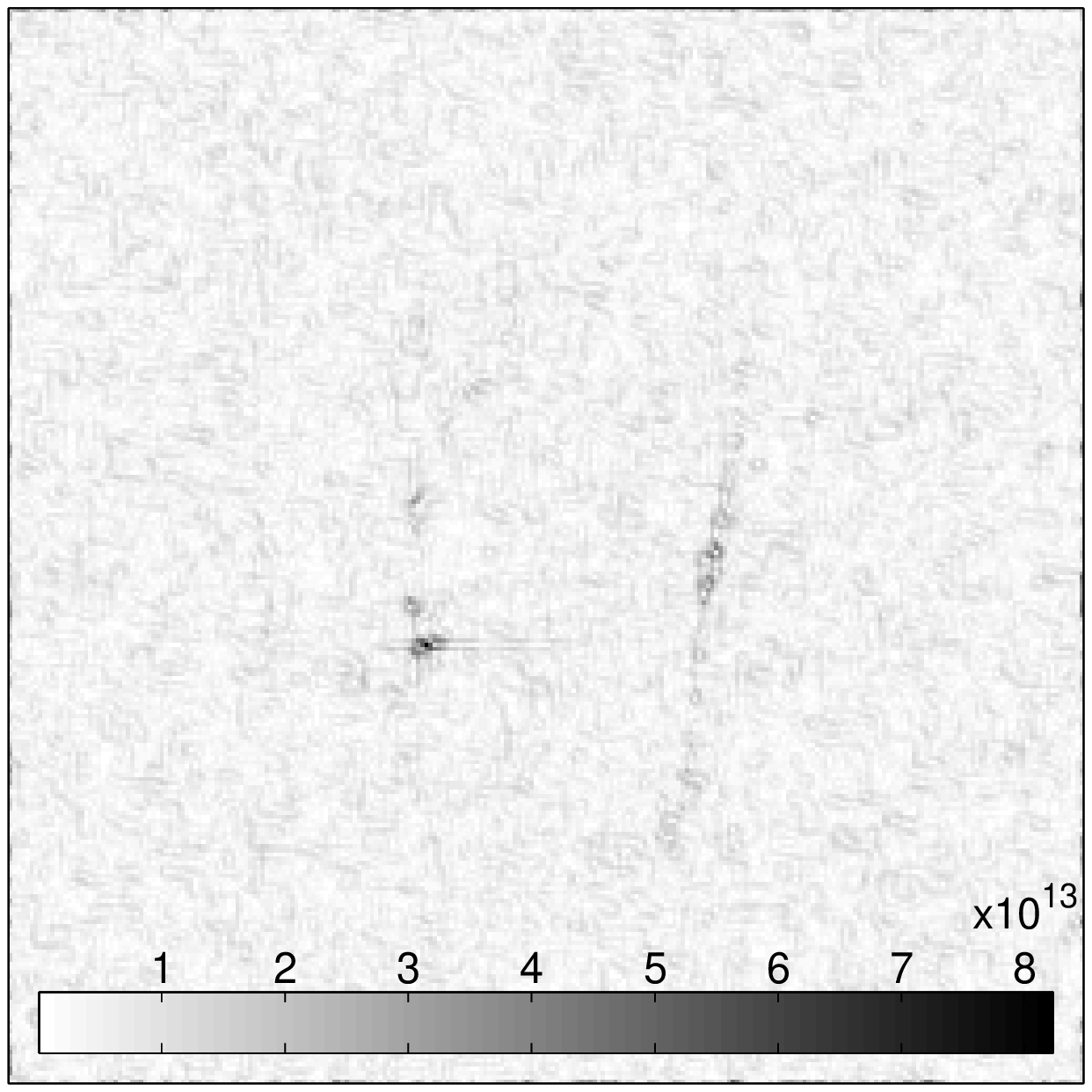}
\hspace{0.2cm}
\includegraphics[height=4cm,keepaspectratio]{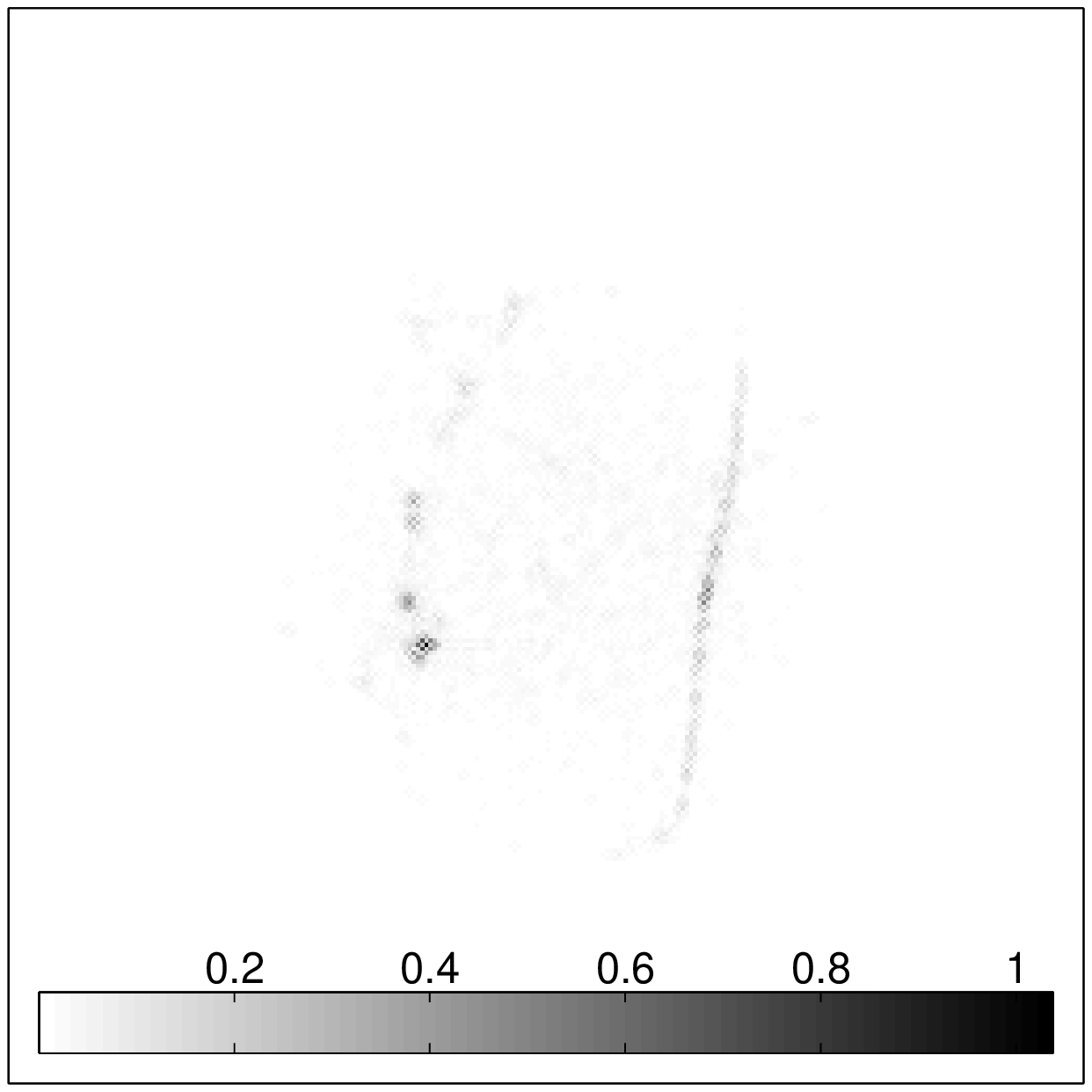}
\hspace{0.2cm}
\includegraphics[height=4cm,keepaspectratio]{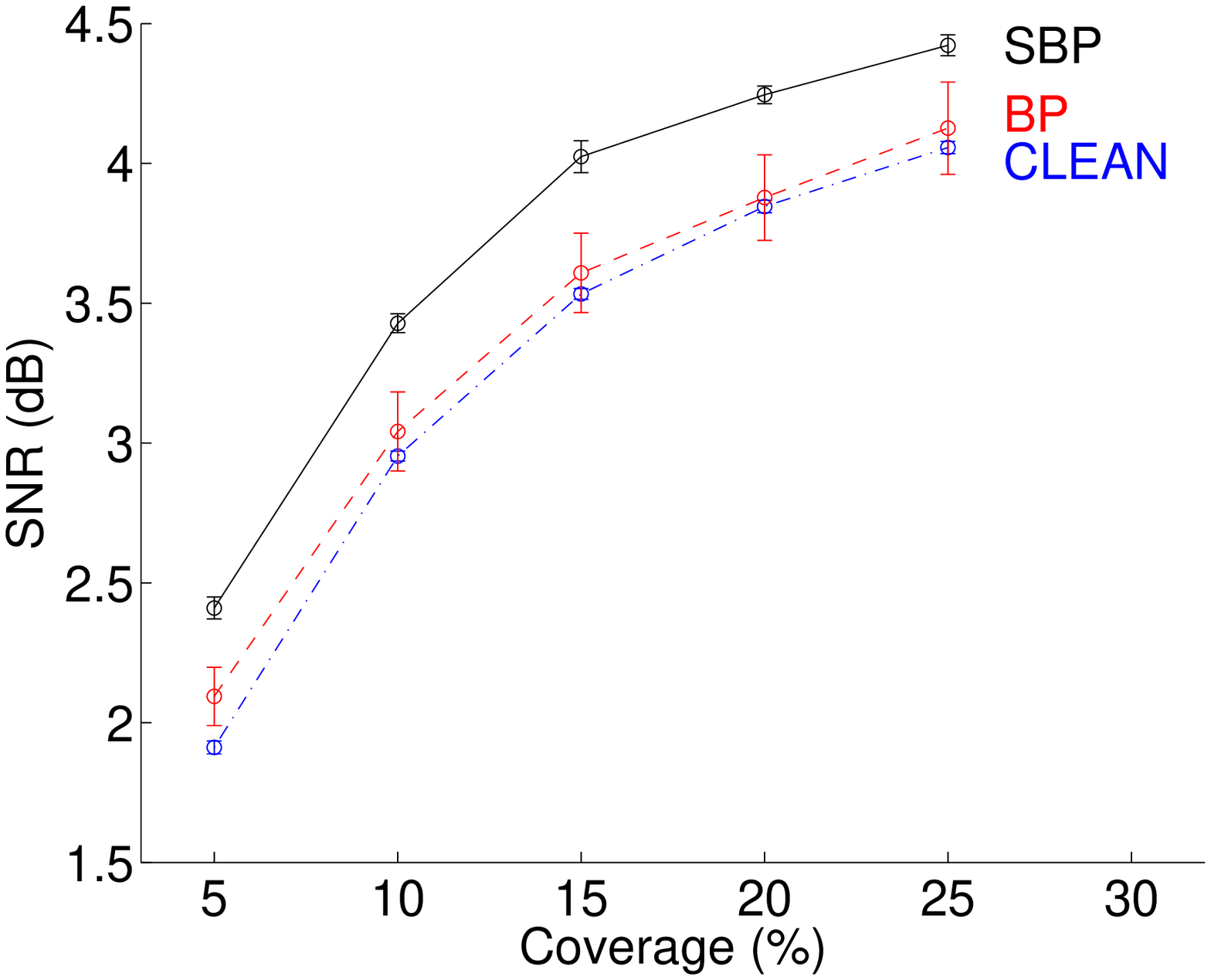}
\end{center}
\caption{\label{fig:all}Top panels: compact object intensity field in some arbitrary intensity units. The original signal multiplied by the illumination function $\bar{x}$ is reported (left), as well as the dirty image $\bar{x}^{(d)}$ (center left) and the BP+ reconstruction of $\bar{x}$ (center right), for the interferometric configuration $c=2$. The graph of the mean SNR with $1\sigma$ error bars over 30 simulations is also reported for the CLEAN, BP, and BP+ reconstructions of $\bar{x}$ as a function of the Fourier coverage identifying the interferometric configurations (extreme right). Bottom panels: string signal in the CMB in $\mu\textnormal{K}$. The magnitude of the gradient of the original signal $x$ re-multiplied by the illumination function is reported (left), as well as the dirty image $\bar{x}^{(d)}$ (center left) and the $\textnormal{SBP}_{\epsilon}$ reconstruction of $x$ re-multiplied by the illumination function (center right), for the interferometric configuration $c=2$. The graph of the mean SNR with $1\sigma$ error bars over 30 simulations is also reported for the CLEAN reconstruction, and for the $\textnormal{BP}_{\epsilon}$ and $\textnormal{SBP}_{\epsilon}$ reconstructions re-multiplied by the illumination function, as a function of the Fourier coverage identifying the interferometric configurations (extreme right).}
\end{figure*}We consider two kinds of astrophysical signals $I$ that are sparse
in some basis, and for which specific prior information is available.
For each kind of signal, $30$ simulations are considered. Observations
of both kinds of signals are simulated for five hypothetical radio
interferometers unaffected by instrumental noise, assuming that the
conditions under which relation (\ref{ri3}) holds are satisfied.
The field of view observed on the celestial sphere by the interferometers
is limited by a Gaussian illumination function $A$ with a full width
at half maximum (FWHM) of $40$ arcminutes of angular opening. The
original signals considered are defined as sampled images with $N=256\times256$
pixels on a total field of view of $1.8^{\circ}\times1.8^{\circ}$.

The first kind of signal consists of a compact object intensity field
in which the astrophysical objects are represented as a superposition
of elongated Gaussians of various scales in some arbitrary intensity
units. The important specific prior information in this case is the
positivity of the signal. The second kind of signal is of particular
interest for cosmology. It consists of temperature steps in $\mu\textnormal{K}$
induced by topological defects such as cosmic strings in the zero-mean
perturbations of the CMB. The string network of interest can be mapped
as the magnitude of the gradient of the string signal itself. The
essential specific prior information in this case resides in the fact
that the statistical distribution of a string signal may be well modelled
in wavelet space. One simulation of a compact object intensity field
and the magnitude of the gradient of one simulation of a string signal
are represented in Figure \ref{fig:all}, after multiplication by
the illumination function.

As discussed already, we assume uniform random selections of visibilities.
The five interferometers considered identified by an index $c$ with
$1\leq c\leq5$ only differ by their Fourier coverage. This coverage
is defined by the $m/2$ randomly distributed frequencies probed in
one half of the Fourier plane, corresponding to $m$ real Fourier
coefficients as: $m/N=5c/100$. For each configuration, the general
inverse problem is the one posed in (\ref{ri4}) with the sensing
matrix $\Phi=\Phi_{_{\!{\rm ri}}}$ defined in (\ref{ri5}) if one
wishes to undo the multiplication by the illumination function and
to recover the original signal $x$. The inverse problem (\ref{eq:ri4'})
applies with the sensing matrix $\Phi=\bar{\Phi}_{_{\!{\rm ri}}}$
defined in (\ref{eq:ri5'}) if one wishes to recover $\bar{x}$.

For each reconstruction, we compare the performance of the Basis Pursuit
approaches enhanced by the inclusion of specific prior signal information
in the minimization problem, with both the standard $\textnormal{BP}_{\epsilon}$
or BP performance, and the CLEAN performance. As the signals considered
are sparse or compressible in some basis, we do not consider any MEM
or WIPE reconstruction, which disregard the sparsity information.
The performance of the algorithms compared is evaluated through the
signal-to-noise ratio (SNR) of the reconstruction for the compact
object intensity field, and through the SNR of the magnitude of the
gradient of the reconstruction for the string signal. The SNR of a
reconstructed signal $\overline{s}$ relative to an original signal
$s$ is technically defined as \begin{equation}
\textnormal{SNR}^{(s,\overline{s})}=-20\log_{10}\frac{\sigma^{(s-\overline{s})}}{\sigma^{(s)}},\label{a1}\end{equation}
where $\sigma^{(s-\overline{s})}$ and $\sigma^{(s)}$ stand for the
sampled standard deviations of the residual signal $s-\overline{s}$
and of the original signal $s$, respectively. It is consequently
measured in decibels (dB).

As far as the computation complexity of the algorithms is concerned,
notice that both CLEAN and the various Basis Pursuit algorithms considered
share the same scaling with $N$ at each iteration. This scaling is
driven by the complexity of the FFT, i.e. $\mathcal{O}(N\log N)$.
The number of iterations required by each algorihm is therefore critical
in a comparison of computation times.

\subsection{Compact object intensity field}

Each simulation of the compact object intensity field consists of
$100$ Gaussians with random positions and orientations, random amplitudes
in the range $[0,1]$ in the chosen intensity units, and random but
small scales identified by standard deviations along each basis direction
in the range $[1,4]$ in number of pixels. Given their structure,
such signals are probably optimally modelled by sparse approximations
in some wavelet basis. But as the maximum possible incoherence with
Fourier space is reached from real space, we chose the sparsity or
compressibility basis to be the Dirac basis, i.e. $\Psi=\mathbb{I}_{N^{1/2}\times N^{1/2}}$.
For further simplification of the problem we consider the inverse
problem (\ref{eq:ri4'}) with the sensing matrix $\bar{\Phi}_{_{\!{\rm ri}}}$,
for reconstruction of the original signal $\bar{x}$ multiplied by
the illumination function.

As no noise is considered, a BP problem is considered in a standard
compressed sensing approach. However, the prior knowledge of the positivity
of the signal also allows one to pose an enhanced BP+ problem as:\begin{equation}
\min_{\bar{x}'\in\mathbb{R}^{N}}\vert\vert\bar{x}'\vert\vert_{1}\textnormal{ subject to }y=\bar{\Phi}_{_{\!{\rm ri}}}\bar{x}'\textnormal{ and }\bar{x}'\geq0.\label{eq:bp+}\end{equation}
Notice that no theoretical recovery result was yet provided for such
a problem in the described framework of compressed sensing. But the
performance of this approach for the problem considered is assessed
on the basis of the simulations. The positivity prior is easily incorporated
into a convex optimization solver based on proximal operator theory
\citep{moreau62}. The Douglas-Rachford splitting method \citep{combettes07}
guarantees that such an additional convex constraint is inserted naturally
in an efficient iterative procedure finding the global minimum of
the BP+ problem. For simplicity, the stopping criterion of the iterative
process is here set in terms of the number of iterations: $10^{4}$.

The BP+ reconstruction of the original signal $\bar{x}$ reported
in Figure \ref{fig:all} is also represented in the figure for the
configuration $c=2$. For comparison, the dirty image $\bar{x}^{(d)}$
used in CLEAN and obtained by simple application of the adjoint sensing
matrix $\bar{\Phi}_{_{\!{\rm ri}}}^{\dagger}$ to the observed visibilities
is also represented. The mean SNR and corresponding one standard deviation
($1\sigma$) error bars over the $30$ simulations are reported in
Figure \ref{fig:all} for the CLEAN reconstruction of $\bar{x}$ with
$\gamma=0.1$, and for the BP and BP+ reconstructions of $\bar{x}$,
as a function of the Fourier coverage identifying the interferometric
configurations. All obviously compare very favorably relative to the
SNR of $\bar{x}^{(d)}$, not reported on the graph. One must acknowledge
the fact that $\textnormal{BP}$ and CLEAN provide relatively similar
qualities of reconstruction. However, the BP reconstruction is actually
achieved much more rapidly than the CLEAN reconstruction, both in
terms of number of iterations and computation time. This highlights
the fact that the BP approach may in general be computationally much
less expensive. The BP+ reconstruction exhibits a significantly better
SNR than the BP and CLEAN reconstructions. The main outcome of this
analysis thus resides in the fact that the inclusion of the positivity
prior on the signal significantly improves reconstruction. For completeness,
let us mention that it was suggested decades ago that CLEAN can be
understood as some approximation of what we called the BP+ approach
\citep{marsh87}.

Notice that the sparsity or compressibility basis is orthonormal and
the error $\vert\vert\bar{x}-\bar{x}^{*}\vert\vert_{2}$ in the BP
reconstruction $\bar{x}^{*}$ of $\bar{x}$ is theoretically bounded
by (\ref{csp5}) with $\epsilon=0$. Assuming saturation of this bound,
the SNR of the BP reconstruction allows the estimation of the maximum
sparsity $K$ of the best sparse approximation $\bar{x}_{K}$ of $\bar{x}$.
Preliminary analysis from the mean SNR of reconstructions over the
simulations considered suggests that ratios $m/K\simeq5$ hold for
each of the values of $m$ associated with the five interferometric
configurations probed. This result appears to be in full coherence
with the accepted empirical ratios quoted above \citep{lustig07}.

\subsection{String signal in the CMB}

The CMB signal as a whole is a realization of a statistical process.
In our setting, the zero-mean temperature perturbations considered
in $\mu\textnormal{K}$ may be modelled as a linear superposition
of the non-Gaussian string signal $x$ made up of steps and of a Gaussian
component $g$ seen as noise. The power spectrum of this astrophysical
noise is set by the concordance cosmological model. We only include
here the so-called primary CMB anisotropies \citep{hammond08}. The
typical number, width and spatial distribution of long strings or
string loops in a given field of view are also all governed by the
concordance cosmological model. Our $30$ simulations of the CMB signal
are built as a superposition of a unique realistic string signal simulation
borrowed from \citet{fraisse08} with $30$ simulations of the Gaussian
correlated noise. The string tension $\rho$, a dimensionless number
related to the mass per unit length of string, is up to some extent
a free parameter of the model. This tension sets the overall amplitude
of the signal and needs to be evaluated from observations. For the
sake of the present analysis, we only study the string signal for
one realistic value $\rho=3.2\times10^{-8}$, which technically fixes
the SNR of the observed string signal buried in the astrophysical
noise. This value is assessed prior to any signal reconstruction,
by fitting the power spectrum of the data to the sum of the power
spectra of the signal and noise on the frequencies probed \citep{hammond08}.
This estimation may be considered as very precise at the tension of
interest and is not to be considered as a significant source of error
in the subsequent reconstruction.

In this context, preliminary analysis of $16$ independent realistic
simulations of a string signal, also from \citet{fraisse08}, allows
one to show that the random process from which the string signal arises
is well modelled by GGD's in wavelet space \citep{hammond08}. We
consider a redundant steerable wavelet basis $\Psi_{_{{\rm \! s}}}$
with $6$ scales $j$ ($1\leq j\leq6$) including low pass and high
pass axisymmetric filters, and four intermediate scales defining steerable
wavelets with $6$ basis orientations $q$ ($1\leq q\leq6$) \citep{simoncelli95}.
By statistical isotropy, the GGD priors $\pi_{j}$ for a wavelet coefficient
$\alpha'_{w}$ only depend on the scale: \begin{equation}
\pi_{j}\left(\alpha_{w}\right)\propto\exp\left[-\Big\vert\frac{\alpha_{w}}{\rho u_{j}}\Big\vert^{v_{j}}\right],\label{a2}\end{equation}
where $w$ is to be thought of as a multi-index identifying a coefficient
at given scale $j$, position $i$, and orientation $q$. Assuming
independence of the wavelet coefficients, the total prior probability
distribution of the signal is simply the product of the probability
distributions for each value of $w$, which reads as \begin{equation}
\pi\left(\alpha\right)\propto\exp-\vert\vert\alpha\vert\vert_{{\rm s}},\label{a3}\end{equation}
for a {}``$\textnormal{s}$'' norm \begin{equation}
\vert\vert\alpha\vert\vert_{{\rm s}}\equiv\sum_{w}\Big\vert\frac{\alpha_{w}}{\rho u_{j}}\Big\vert^{v_{j}}.\label{a4}\end{equation}
The exponent parameters $v_{j}$ are called GGD shape parameters and
can be considered as a measure of the compressibility of the underlying
distribution. Values close to $0$ yield very peaked probability distributions
with heavy tails relative to Gaussian distributions, i.e. very compressible
distributions. The list of these values at all scales reads as: $\{v_{1}=0.43,v_{2}=0.39,v_{3}=0.47,v_{4}=0.58,v_{5}=0.76,v_{6}=1.86\}$.
The signal is thus understood as being well modelled by a very compressible
expansion in its wavelet representation and we choose the corresponding
redundant basis as the sparsity or compressibility basis for the inverse
problem: $\Psi=\Psi_{_{\!{\rm s}}}$. The list values of the GGD scale
parameters $u_{j}$ identifying the variances of the distributions
at all scales reads as: $\{u_{1}=8.9\times10^{-3},u_{2}=2.8\times10^{-3},u_{3}=2.2\times10^{-2},u_{4}=0.15,u_{5}=0.95,u_{6}=57\}$.
In full generality we consider the general inverse problem (\ref{ri4})
with the sensing matrix $\Phi_{_{{\rm \! ri}}}$, for reconstruction
of the original signal $x$ non-multiplied by the illumination function.

Even in the absence of instrumental noise the measured visibilities
thus follow from (\ref{csp2}) with a noise term \begin{equation}
n=\Phi_{_{\!{\rm ri}}}g,\label{a5}\end{equation}
representing values of the Fourier transform of the astrophysical
noise $g$ multiplied by the illumination function. Discarding the
very local correlations in the Fourier plane introduced by the illumination
function, one may consider that the measurements are independent and
affected by independent Gaussian noise realizations. The corresponding
noise variance $\sigma_{r}^{2}$ on $y_{r}$ with $1\leq r\leq m$,
is thus identified from the values of the known power spectrum of
$g$.

A whitening matrix $W_{_{\!{\rm cmb}}}\in\mathbb{R}^{m\times m}=\{(W_{_{\!{\rm cmb}}})_{rr'}=\sigma_{r}^{-1}\delta_{rr'}\}_{1\leq r,r'\leq m}$
is introduced on the measured visibilities $y$, so that the corresponding
visibilities $\tilde{y}=W_{_{\!{\rm cmb}}}y$ are affected by independent
and identically distributed noise, as required to pose a $\textnormal{BP}_{\epsilon}$
problem. This operation corresponds to a matched filtering in the
absence of which any hope of good reconstruction is vain. A $\textnormal{BP}_{\epsilon}$
problem is thus considered after estimation of $\rho$. However, the
prior statistical knowledge on the signal also allows one to pose
an enhanced Statistical Basis Pursuit denoise ($\textnormal{SBP}_{\epsilon}$)
problem. It is defined as the minimization of the negative logarithm
of the specific prior on the signal, i.e. the $\textnormal{s}$ norm
of the vector of its wavelet coefficients, under the measurement constraint:\begin{equation}
\min_{\alpha'\in\mathbb{R}^{T}}\vert\vert\alpha'\vert\vert_{{\rm s}}\textnormal{ subject to }\vert\vert\tilde{y}-W_{_{\!{\rm cmb}}}\Phi_{_{\!{\rm ri}}}\Psi_{_{{\rm \! s}}}\alpha'\vert\vert_{2}\leq\epsilon.\label{eq:sbpepsilon}\end{equation}
Notice that the $\textnormal{s}$ norm is similar but still more general
than a single $\ell_{p}$ norm and no theoretical recovery result
was yet provided for such a problem in the framework of compressed
sensing. Again, the performance of this approach for the problem considered
is assessed on the basis of the simulations. Most shape parameters
$v_{j}$ are smaller than $1$, which implies that the norm defined
is not convex. We thus reconstruct the signal through the re-weighted
$\ell_{1}$ norm minimization described above \citep{candes08a}.
In this regard, we use the SPGL1 toolbox \citep{vandenBerg08}%
\footnote{http://www.cs.ubc.ca/labs/scl/spgl1/%
}. The value of $\epsilon^{2}$ in the $\textnormal{BP}_{\epsilon}$
and $\textnormal{SBP}_{\epsilon}$ problems is taken to be around
the $99^{\textnormal{th}}$ percentile of the $\chi^{2}$ with $m$
degrees of freedom governing the noise level estimator. This value
also serves as the stopping criterion for the CLEAN reconstruction.

The magnitude of the gradient of the $\textnormal{SBP}_{\epsilon}$
reconstruction of the original signal $x$ reported in Figure \ref{fig:all}
is also represented in the figure for the configuration $c=2$, after
re-multiplication by the illumination function which sets the field
of view of interest. For comparison, the magnitude of the gradient
of the dirty image $\bar{x}^{(d)}$ used in CLEAN and obtained by
simple application of the adjoint sensing matrix $\bar{\Phi}_{_{\!{\rm ri}}}^{\dagger}$
to the observed visibilities is also represented. 

The mean SNR and corresponding one standard deviation ($1\sigma$)
error bars over the $30$ simulations are reported in Figure \ref{fig:all}
for the CLEAN reconstruction with $\gamma=0.1$, and for the $\textnormal{BP}_{\epsilon}$
and $\textnormal{SBP}_{\epsilon}$ reconstructions re-multiplied by
the illumination function, as a function of the Fourier coverage identifying
the interferometric configurations. All obviously compare very favorably
relative to the SNR of $\bar{x}^{(d)}$, not reported on the graph.
One must still acknowledge the fact that $\textnormal{BP}_{\epsilon}$
and CLEAN provide relatively similar qualities of reconstruction.
The $\textnormal{BP}_{\epsilon}$ reconstruction is achieved much
more rapidly than the CLEAN reconstruction, highlighting the fact
that the $\textnormal{BP}_{\epsilon}$ approach may in general be
computationally much less expensive. The $\textnormal{SBP}_{\epsilon}$
reconstruction exhibits a significantly better SNR than the BP and
CLEAN reconstructions.

Let us acknowledge the fact that the re-weighted $\ell_{1}$ norm
minimization of the $\textnormal{SBP}_{\epsilon}$ approach proceeds
by successive iterations of $\ell_{1}$ norm minimization. This unavoidably
significantly increases the computation time for reconstruction relative
to the single $\ell_{1}$ norm minimization of the $\textnormal{BP}_{\epsilon}$
approach. Relying on the idea that the coefficients of the low pass
filter do not significantly participate to the identification of the
string network itself, our implementation of $\textnormal{SBP}_{\epsilon}$
does not perform any re-weighting at the scale $j=6$, where $v_{6}=1$
was thus assumed. This restriction allows one to keep $\textnormal{SBP}_{\epsilon}$
computation times similar to those of CLEAN. Let us notice however
that an even better SNR is obtained by correct re-weighting at $j=6$,
albeit at the cost of a prohibitive increase in computation time.

The main outcome of the analysis is twofold. Firstly, the presence
of a whitening operation is essential when correlated noise is considered.
Secondly, the inclusion of the prior statistical knowledge on the
signal also significantly improves reconstruction.

\section{Conclusion}

\label{sec:Conclusion}Compressed sensing offers a new framework for
image reconstruction in radio interferometry. In this context, the
inverse problem for image reconstruction from incomplete and noisy
Fourier measurements is regularized by the definition of global minimization
problems in which a generic sparsity or compressibility prior is explicitly
imposed. These problems are solved through convex optimization. The
versatility of this scheme also allows inclusion of specific prior
information on the signal under scrutiny in the minimization problems.
We studied reconstruction performances on simulations of an intensity
field of compact astrophysical objects and of a signal induced by
cosmic strings in the CMB temperature field, observed with very generic
interferometric configurations. The $\textnormal{BP}_{\epsilon}$
technique provides similar reconstruction performances as the standard
matching pursuit algorithm CLEAN. The inclusion of specific prior
information significantly improves the quality of reconstruction.

Further work by the authors along these lines is in preparation. In
particular, a more complete analysis is being performed to estimate
the lowest string tension down to which compressed sensing imaging
techniques can reconstruct a string signal in the CMB, in more realistic
noise and Fourier coverage conditions. In this case, given the compressibility
of the magnitude of the gradient of the string signal itself, TV norm
minimization also represents an interesting alternative to the $\textnormal{SBP}_{\epsilon}$
problem proposed here.

\section*{Acknowledgments}

The authors wish to thank A. A. Fraisse, C. Ringeval, D. N. Spergel,
and F. R. Bouchet for kindly providing string signal simulations,
as well as M. J. Fadili for discussions on optimization by proximal
methods. The authors also thank the reviewer T. J. Cornwell for his
valuable comments. The work of Y. W. was funded by the Swiss National
Science Foundation (SNF) under contract No. 200020-113353. Y. W. and
L. J. are Postdoctoral Researchers of the Belgian National Science
Foundation (F.R.S.-FNRS).

\label{lastpage}

\end{document}